\documentclass[a4paper,12pt]{article}
\usepackage{graphicx}
\usepackage{amsmath,amssymb}

\usepackage{bm}
\usepackage{verbatim}
\usepackage{amsmath,amssymb}
\usepackage{relsize,exscale}
\usepackage{color}
\begin{document}

\title{ Lagrangian form of  Schr\"odinger equation}

\author{D. Arsenovi\' c$^1$, N. Buri\' c $^1${\thanks {buric@ipb.ac.rs}}, D.M. Davidovi\' c$^2$\\
 and S. Prvanovi\' c$^1$\\
1. Institute of Physics, University of Belgrade, \\PO Box 68, 11000 Belgrade, Serbia.\\
2. Vinca Institute of Nuclear Sciences, University of Belgrade, Serbia.
}
\maketitle
\begin{abstract}
Lagrangian formulation of quantum mechanical Schr\"odinger equation is developed in general and illustrated in the eigenbasis of the Hamiltonian and in the coordinate representation. The Lagrangian formulation of physically plausible quantum system results in a well defined second order equation on a real vector space. The Klein-Gordon equation for a real field is shown to be the Lagrangian form of the corresponding Schr\"odinger equation.

\end{abstract}


\section {Introduction}

Schr\" odinger evolution equation of quantum mechanics \cite{Dirac} is of the first order in time and is formulated on the complex  Hilbert space
of the system. In these respects, it is similar to the evolution equation of a Hamiltonian dynamical system\cite{Arnold,Del1}. The latter have
an equivalent Lagrangian formulation in terms of equations of the second order in time. On the other hand, something like a Lagrangian formulation
 of the abstract Schr\" odinger equation is usually not considered. The question of a second order evolution equation equivalent to the Schr\"odinger equation in coordinate representation was considered for the first time by Schr\" odinger himself \cite{Schr}. However, he did not consider the problem as the one of relation between Hamiltonian (Schr\" odinger) and Lagrange frameworks. An interesting analysis of this problem, again formulated in the coordinate representation, appeared in \cite{Del2}. Construction of an appropriate Lagrangian formulation of the quantum Schr\" odinger evolution in the general case is the task of this paper. This is quite different from the standard application of the action functional in the Lagrangian (or Hamiltonian ) form of the functional integral representation of the quantum propagator \cite{Dirac1,Fayman}.

 \begin{figure}[t]
\centering
\includegraphics[width=0.8\textwidth]{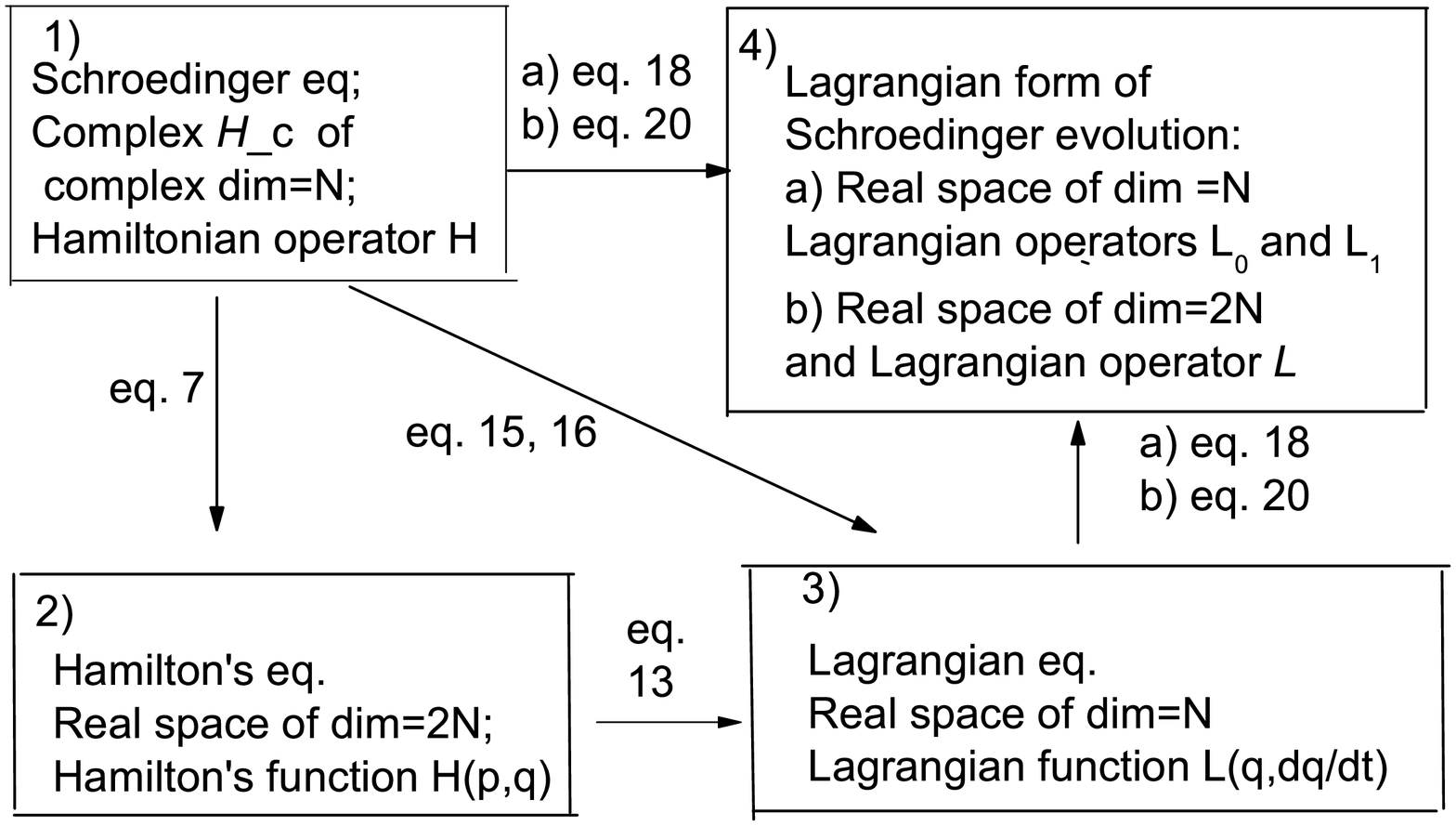}
\caption{Illustrates relations between various formulations of quantum Schr\"odinger evolution.
\label{Fig1}}
\end{figure}

 Our strategy and the main result could be schematically summarized as in figure 1. We shall start from the quantum Schr\" odinger equation on an abstract complex Hilbert space, which we consider as given. This is the box 1 in fig.1. It is well known that the Schr\" odinger equation is equivalent to the appropriate Hamiltonian dynamical system on the corresponding phase space \cite{Kible, Heslot, Brody1,Ashtecar}. This is the box 2 in fig.1.
  In fact, the geometric Hamiltonian form of the Schr\" odinger evolution could be extended to an equivalent formulation of the quantum mechanics
  \cite{Ashtecar}, and is quite suitable for treatment of problems like
   nonlinear constraints\cite{JaAnnphys, Brody2}, imbedding of classical into quantum mechanics \cite{usPRA1,usPRA2} and hybrid theories \cite{Elze,usPRA3,usPRA4}. We shall apply the standard transition procedure from Hamiltonian to Lagrangian formulation in order to obtain the Lagrangian form of the Hamiltonian formulation of the Schr\" odinger equation. The result is the box 3 in fig.1. The final step is the transition
    from the Lagrangian dynamical system in box 3 to the Lagrangian form of the Schr\" odinger equation, which turns out to be a second order equation on a real Hilbert space. This is the box 4 in fig. 1. The transitions between boxes, represented by the arrows in fig. 1, are explained in detail, and in the general case, in the next section.

     Of course,  the Lagrangian form,  when it exists, is equivalent to the Hamiltonian Schr\" odinger equation. Therefore,
     our main motivation for the derivation of the Lagrangian form of the general (linear) Schr\" odinger equation is purely formal. One further motivation is that the Lagrangian form of the Schr\"odinger equation is more suitable for Lorentz invariant models. In fact, we shall see that the Klein-Gordon equation  \cite{KG} for the relativistic state vector is noting else but the Lagrangian formulation of the Schr\" odinger equation with the corresponding Hamiltonian.
     Also, symmetries are more naturally considered as
   morphisms of the Lagrangian form and not as time evolution, while the Hamiltonian formulation, i.e. the Schr\" odinger equation, on the other hand, offers the framework in which the time evolution is an automorphism of the relevant symplectic structure. However, one should bare on mind that the canonical coordinates and momenta in the Hamiltonian and velocities in the Lagrangian formulations of the Schr\" odinger equation are in fact the coefficients in a basis expansion of the quantum state vector.

  The presentation in section 2 is general and formal. The conditions for existence of the expression of the canonical momenta  in terms of velocities will be briefly discussed in the general case near the end of section 2. The possible problems will be further illustrated and discussed in section 3,
  together with the presentation of particular examples obtained by rewriting the general formalism  in particular bases. In this section we also
   treat the Klein-Gordon equation  as an example of the Lagrangian formulation of the appropriate Schr\"odinger equation.
   Summary is given in section 4.

\section {Derivation of the Lagrangian equations}

 The Schr\" odinger equation
\begin{equation}
i\hbar\frac{d|\psi\rangle}{dt}=\hat H|\psi\rangle
\end{equation}
on a complex Hilbert space ${\cal H}_c$ of finite or infinite complex dimension $N$,   is equivalent to a Hamiltonian dynamical system on an appropriate real, symplectic manifold of dimension $2N$. This Hamiltonian formulation
 of the Schr\" odinger evolution equation is well known ( excellent comprehensive reviews are \cite{Ashtecar, Brody1}). Nevertheless, we need to recapitulate the notation and write down the basic definitions and formulas in their most elementary form. Derivation of the Hamiltonian form of the equation (1) ends with the equations (10) and the rest of this section is devoted to
    the derivation  of the Lagrangian form of the Schr\" odinger evolution.

     A vector $|\psi\rangle\in {\cal H}_c$ in an abstract (arbitrary) basis $\{|n\rangle\}$ is written as
 \begin{equation}
  |\psi\rangle=\sum_n z_n|n\rangle=\sum_n(q_n+ip_n)|n\rangle
  \end{equation}
The matrix elements
 \begin{equation}
 H_{mn}=\langle m|\hat H|n\rangle\equiv H_{mn}^R+iH_{mn}^I
  \end{equation}
  satisfy $ H_{mn}^R= H_{nm}^R,\>  H_{mn}^I=- H_{nm}^I$ since $\hat H$ is a Hermitian operator.
   It is not necessary for our main line of reasoning, but it is nevertheless instructive, to write down explicitly the abstract matrix form of the Schro\" odinger equation (1) in terms of real $2N$ dimensional vector
\begin{equation}
|\phi\rangle=(q_1,\dots q_n,p_1\dots p_n)^T\equiv (q,p)^T
\end{equation}
The Schr\" odinger equation (1) in terms of the $2N$ real vector $|\phi\rangle$ in the abstract basis reads
\begin{equation}
\frac{d|\phi\rangle}{dt}=\frac{1}{ \hbar}\begin{pmatrix}\hat H^I&\hat H^R\cr
-\hat H^R&\hat H^I\end{pmatrix}|\phi\rangle,
\end{equation}
where ${\hat H^I, \hat H^R}$ denote $N\times N$ matrices introduced in (3). The Schr\" odinfer equation (1) and the representations in the abstract
 basis of ${\cal H}_c$ are the content of the box 1, and are considered as given.

 The function
\begin{equation}
H(q,p)=\frac{1}{2\hbar}\langle \psi|\hat H|\psi\rangle
\end{equation}
where $q_n,p_n,|\psi\rangle$ are as in (2) is a quadratic form given explicitly by
\begin{equation}
 H(q,p)=\frac{1}{2\hbar}\sum_{nm}\left( q_nq_m H^R_{nm}+p_np_mH^R_{nm}-2q_np_mH_{nm}^I\right ).
 \end{equation}
 Let us remark that the function $H(q,p)$ can also be written as $\langle \phi|\hat H^C|\phi\rangle$ where $|\phi\rangle$ is given by (4) but the
   representation is not unique, i.e. the expression (7) is given by $\langle \phi|\hat H^C|\phi\rangle$ where
\begin{equation}
\hat H^c=\frac{1}{2 \hbar}\begin{pmatrix}\hat H^R&-\hat H^I\cr
\hat H^I&\hat H^R\end{pmatrix}
\end{equation}
or
\begin{equation}
\hat H^c=\frac{1}{2 \hbar}\begin{pmatrix}\hat H^R&-2\hat H^I\cr
0&\hat H^R\end{pmatrix}.
\end{equation}
or other similar matrces.

It is easily checked that the Hamilton equations with the Hamilton's function $H(q,p)$ (7):
\begin{equation}
\dot q_n=\frac{\partial H(q,p)}{\partial p_n},\quad \dot p_n=-\frac{\partial H(q,p)}{\partial q_n}
\end{equation}
reproduce the real form of the Schr\" odinger equation (5). The Hamiltonian function (6) and (7) and the corresponding Hamilton's equations are the
 content of the box 2, and the arrow from box 1 to box 2 is given explicitly by the formula (7).

In order to derive the Lagrangian form of the Hamilton's equations (10) we need to express the canonical momenta $p_n$ in terms of
 the velocities $\dot q_n$. This is done in the standard way, and using the explicit form of the equations (10). In fact, from (10a)
 \begin{equation}
 \sum_m H_{nm}^Rp_m= \hbar \dot q_n-\sum_m H_{nm}^Iq_m
 \end{equation}
Therefore
\begin{equation}
p_k=\sum_n \left ((H^R)^{-1}\right )_{kn}(\hbar \dot q_n-\sum_{m} H_{nm}^I q_m),
\end{equation}
 or in a more compact form $p=(\hat H^R)^{-1}(\hbar \dot q-\hat H^I q)$. At this point we do not care about the existence of the inverse operator $(\hat H_R)^{-1}$.
  The relation (12) is substituted into the definition of the Lagrange function
   $L=\sum_l p_l \dot q_l-H(q,p)$ to obtain, after some manipulation
   \begin{eqnarray}
   L(q,\dot q)&=&\frac{\hbar}{2}\sum_{n,m} \dot q_n \left( (H^R)^{-1}  \right)_{nm}\dot q_m+\sum_{n,m,k} q_n  H^I_{nk}\left ((H^R)^{-1}  \right)_{km}\dot q_m\nonumber\\
   &-&\frac{1}{2\hbar}\sum_{n,m}q_n\left [\sum_{k,j}H_{n,k}^I((H^R)^{-1})_{k,j} H_{j,m}^I+H_{n,m}^R\right ]q_m
   \end{eqnarray}
   The previous formula for the Lagrangian function $L(q,\dot q)$ in terms of the Hamiltonian $\hat H$ could be rewritten in a perhaps more systematic form
   \begin{equation}
   L(q,\dot q)=\sum_{nm}\dot q_n L^{\dot q\dot q}_{nm} \dot q_m+\sum_{nm} q_n L^{ q\dot q}_{nm} \dot q_m+\sum_{nm} q_n L^{ q q}_{nm} q_m
\end{equation}
where $L^{\dot q\dot q},L^{q\dot q},L^{qq}$ are  given in terms of $\hat H^R, (\hat H^R)^{-1}, \hat H^I$ in the following compact form
\begin{equation}
 L^{\dot q\dot q}=\frac{\hbar}{2}(\hat H^R)^{-1},\> L^{ q\dot q}=\hat H^I(\hat H^R)^{-1},\> L^{ q q}=\frac {-1}{2\hbar}[H^I(\hat H^R)^{-1} \hat H^I+\hat H^R].
 \end{equation}
 The Lagrange function (14) is uniquely related to the original Hamiltonian operator in (1).
 The corresponding Lagrangian equations are
 \begin{eqnarray}
 \ddot q_m&=&\frac{-1}{2}\sum_{n,i}(L^{\dot q\dot q}_{m,i})^{-1} L^{q\dot q}_{n,i} \dot q_n+\frac{1}{2}\sum_{n,i}(L^{\dot q\dot q}_{m,i})^{-1} L^{q\dot q}_{i,n} \dot q_n\nonumber\\
 &+&\sum_{n,i}(L^{\dot q\dot q}_{m,i})^{-1} L^{q q}_{i,n} q_n,
 \end{eqnarray}
 where $L^{\dot q\dot q}, L^{ q\dot q}, L^{ q q}$ are given by (15).

   The Lagrangian (14) with transition formulas (15) is the content of the box 3 and the arrow from the box 1 to the box 3.

 Let us remark that the Lagrangian (14) can be written in the form resembling the equation (6) by introducing 2N-dimensional vector $|\kappa\rangle =(q,\dot q)^T$. Then
 \begin{equation}
 L(q,\dot q)=\langle \kappa| \begin{pmatrix}\frac {-1}{2\hbar}[H^I(\hat H^R)^{-1} \hat H^I+\hat H^R]&\frac{1}{2}\hat H^I (\hat H^R)^{-1}\cr
\frac{-1}{2}(\hat H^R)^{-1}\hat H^I&\frac{\hbar}{2}(\hat H^R)^{-1}\end{pmatrix}|\kappa\rangle
  \end{equation}

 In order to go from the box 3 to the box 4 we introduce operators $\hat L_0$ and $\hat L_1$, acting on a real N-dimensional space, which are defined in terms of $\hat H^R,\hat H^I, (\hat H^R)^{-1}$ (eq. (3)) as follows
  \begin{equation}
  \hat L_0=\frac{-1}{\hbar^2}[\hat H^R \hat H^I(\hat H^R)^{-1}\hat H^I+(\hat H^R)^2],\quad \hat L_1=\frac{1}{\hbar}[\hat H^I+\hat H^R\hat H^I(\hat H^R)^{-1}].
  \end{equation}
  The Lagrangian equations (16) are written as
  \begin{equation}
  \frac{d^2| q\rangle}{dt^2}= \hat L_1\frac{d| q\rangle}{dt}+\hat L_0|q\rangle
  \end{equation}
  where $|q\rangle \in {\cal H}_R$. This is the content of the box 4. and represents the Lagrangian form of the quantum Schroedinger equation on the
   complex space ${\cal H}_c$.

   Suppose that the operators $\hat L_0$ and $\hat L_1$ satisfying (18) exist for a given operator $\hat H$ in (1). Then any orbit $|q(t)\rangle$ of (19) in ${\cal H}_R^N$
    can be used to construct the corresponding orbit of the Schr\"odinger equation on ${\cal H}_c^N$, and the Schr\"odinger orbit is given by $|\psi(t)\rangle=|q(t)\rangle+i|p(t)\rangle$, where $|p(t)\rangle$ is given in terms of $|q(t)\rangle$ and
     $|\dot q(t)\rangle$ by the formula (12). Also, the initial conditions for the Lagrangian formulation $q(t_0),\dot q_(t_0)$ are related to the initial
 state of the Hamiltonian equations or of the Schr\" odinger equation essentially by the equation (12).
      Thus, the crucial question for the construction of the Lagrangian formulation is the existence of
    the operators $\hat L_0, \> \hat L_1 $, which is essentially the question of the existence of the inverse operator $(\hat H^R)^{-1}$.
     The Hamiltonian operator $\hat H$ in (1) uniquely determines the operators $\hat H^R$ and $\hat H^I$. However, a singular choice of basis
      in ${\cal H}_c$ might imply that the operator $\hat H^R$ (or $\hat H^I$) is represented by zero and $(H_{nm}^R)^{-1}$ does not exist (see example in 3.1). However,
       in a typical basis $ \hat H^R$ is represented by a nonzero operator. The existence of $(\hat H^R)^{-1}$
 then depends on the physical problem. If $(\hat H^R)^{-1}$, for the system with the Hilbert space ${\cal H}_c$ does not exist, then redefinition of
 the system by restriction on an appropriate subspace of ${\cal H}_c$ would lead to a well defined  $(\hat H^R)^{-1}$. Alternatively, in terms of the Hamiltonian and the corresponding Lagrangian formulations nonexistence of $(\hat H^R)^{-1}$  corresponds to constrained Hamiltonian and singular Lagrangian systems,
  and the relation between the two formalisms is treated by the appropriate methods \cite{Del1}.

   The Lagrangian system (18) on $N$ dimensional ${\cal H}_R$ can be trivially written in the form of an evolution equations on real 2N dimensional real space:
   \begin{equation}
   \frac{d|\kappa\rangle}{dt}={\cal L}|\kappa\rangle
   \end{equation}
   where
   \begin{equation}
 {\cal L}=  \begin{pmatrix}0&1\cr
\hat L_0&\hat L_1\end{pmatrix},
   \end{equation}
and $|\kappa\rangle\in {\cal H}_R^{2N}$ was defined just before the equation (17).

\section{Examples and discussion}

In this section we present a series of examples, in the order of increasing complexity, which are aimed to illustrate
 the problems that might occur in the construction of the Lagrangian formulation.

\subsection{Discrete finite basis}
The first example is rather trivial and artificial, but points out that problems in the construction of the Lagrangian formulation might be
 apparent and induced by  specific choice of a singular basis.
Consider the simplest possible quantum system of a single $1/2$ spin with the Hamiltonian $\hat H=\hat\sigma_y$. In the eigenbasis of the $\hat\sigma_z$
the Hamiltonian is represented by
\begin{equation}
  {\hat H}_{mn}=  \begin{pmatrix} 0&-i\cr
i&0\end{pmatrix}
   \end{equation}
The real part of the Hamiltonian is zero, there is no $(H_{nm}^R)^{-1}$
and the Hamilton's function is
\begin{equation}
H(q,p)=\frac{1}{\hbar}(q_1p_2-q_2p_1)
\end{equation}
 and there apparently is no Lagrangian formulation. However, by an arbitrary small change of  basis the real part becomes nonzero.  In  the eigenbasis of $\hat \sigma_y$ the imaginary part becomes zero and the real part of the Hamiltonian becomes
 \begin{equation}
  {\hat H}_{mn}=  \begin{pmatrix} 1&0\cr
0&-1\end{pmatrix}
   \end{equation}
   The Lagrangian formulation is now constructed as in the next example.

\subsection {Discrete energy eigenbasis}

Derivation and the formulas of the Lagrangian formalism acquire specially simple form in the eigenbasis of the Hamiltonian $\hat H$ with a discrete,
 say non-degenerate, spectrum. The Hamilton's function is of the form corresponding to a system of linear oscillators
 \begin{equation}
 H(q,p)=\sum_m \frac{e_m}{2\hbar}(q_m^2+p_m^2),
\end{equation}
Thus, $H_{nm}=H_{nm}^R = \delta_{nm} e_m$, where $e_m$ are real and $H^I=0$.
 The momenta $p_m$ are related to the velocities $\dot q_m$ by $p_m=\hbar\dot q_m/e_m$, which is well defined if $\hat H$ has no zero eigenvalue.
Lagrangian function (13) is given by
\begin{equation}
L(q,\dot q)=\sum_m \left(\hbar\frac{\dot q_m^2}{e_m}-\frac{e_mq_m^2}{2\hbar}-\hbar\frac{\dot q_m^2}{2 e_m}\right)
\end{equation}
and the Lagrangian equations are
\begin{equation}
\ddot q_m=-\frac{e_m^2}{\hbar^2} q_m,
\end{equation}
or
\begin{equation}
  \ddot\phi=\hat L_0\phi,
  \end{equation}
where $\hat L_0=-\hat H^2/\hbar^2$ operates on the real space ${\mathbf R}^N$. The only possible obstacle to the existence of the Lagrangian formulation is possible zero eigenvalue of the Hamiltonian. If there is a zero eigenvalue then the restriction of the Schr\" odinger equation
 onto the orthogonal complement of the zero eigenspace generates the Hamiltonian system with the canonical pair $q_0,p_0$ that does
  not appear in the Hamilton's function (25), and for which there is the corresponding Lagrangian formulation on ${\mathbf R}^{N-1}$.

From these two examples we see that the construction of the Lagrangian formulation might fail if one makes a choice of the singular basis and/or
 includes nonphysical states.

\subsection {Coordinate representation}

 Construction of the Lagrangian formulation for the Schr\"odinger equation in the coordinate representation
  follows the same steps as in the general case, or can be obtained by applying the general formulas (18) and (19) written in the coordinate basis.
    The only potential problem is, like in the general case, non-existence of the inverse of the real part of the Hamiltonian operator $(\hat H^R)^{-1}$, appearing in (12). This operator typically has non-diagonal elements in almost all bases and, of course, this fact is irrelevant for the question of its existence. In the coordinate
 representation, the non-diagonal character  of $(\hat H^R)^{-1}$  appears as non-locality of the relevant differential operator, and this fact, like its analog in the general case, is irrelevant for the existence of the Lagrangian formulation.

Consider a Hamiltonian of the
 form
 \begin{equation}
 \hat H=\frac{\hat P_x^2}{2m}+V(\hat X).
 \end{equation}
In the representation of the coordinate ${\hat X}$, the evolution equation (1) becomes the  Schr\"odinger linear  partial differential
equation of mathematical physics
\begin{equation}
i\hbar\frac {\partial \psi(t,x)}{\partial t}=-\frac{\hbar^2}{2m}\Delta \psi(t,x)+V(x)\psi(t,x),
\end{equation}
where $\Delta=\frac{\partial^2}{\partial x^2}$.
This equation is equivalent to the Hamiltonian functional equations for the canonical fields $\phi(x),\pi(x)$ introduced by $\psi(t,x)=\phi(t,x)+i\pi(t,x)$. The Hamilton's functional is obtained by applying the general rule
\begin{eqnarray}
H(\phi,\pi)&=&\langle \psi|\hat H|\psi\rangle=\int dx \frac{1}{2\hbar}(\phi-i\pi)(-\frac{\hbar^2}{2m}\Delta+V)(\phi+i\pi)\nonumber\\
&=&\int dx\frac{1}{2\hbar}\left[ -\frac{\hbar^2}{2m}(\phi\Delta\phi+\pi\Delta\pi)+V(\phi^2+\pi^2)\right]
\end{eqnarray}
The Hamilton's functional (31) is traditionally (see for example \cite{Del1, Ashtecar}  further transformed into
\begin{equation}
H(\phi,\pi)=\frac{1}{2\hbar}\int_X dx \left [ \frac{\hbar^2}{2m}((\partial_x\phi(x))^2+(\partial_x\pi(x))^2)+V(x)(\phi(x)^2+\pi(x)^2)\right ],
\end{equation}
but the form (31) is much more suitable for the construction of the Lagrangian formulation.
The  Hamilton's equations corresponding to (31) (or to (32)) are
\begin{eqnarray}
\dot\phi&=&\frac{\delta H}{\delta \pi}=\frac{1}{\hbar}(V-\frac{\hbar^2}{2m}\Delta)\pi\nonumber\\
\dot\pi&=&-\frac{\delta H}{\delta \phi}=\frac{1}{\hbar}(\frac{\hbar^2}{2m}\Delta-V)\phi.
\end{eqnarray}
The generalized momentum $\pi(x)$ is expressed via $\dot\phi(x)$ using the formula (33a) and reads
\begin{equation}
\pi(x)=\hbar (V-\frac{\hbar^2}{2m}\Delta)^{-1}\dot\phi
\end{equation}
Observe that the differential operator $V-\frac{\hbar^2}{2m}\Delta$ maps real functions into real ones and therefore the formula (34) is a special case
 of (12) $p=(\hat H^R)^{-1}\dot q$. Using the equation (34) to replace $\pi$ and (33a) to replace $\Delta\pi$, the Lagrangian functional corresponding to (31) is formed by the standard rule and reads
 \begin{equation}
 L(\phi,\dot\phi)=\int dx \left[\frac{-1}{2\hbar}\phi(V-\frac{\hbar^2}{2m}\Delta)\phi+\frac{\hbar}{2}\dot\phi (V-\frac{\hbar^2}{2m}\Delta)^{-1}\dot \phi\right].
 \end{equation}
 This is just the general formula (13) with $ H_{nm}^I=0$. Observe that the expression $\Delta\pi$ that appears in the Hamiltonian (31) is easily handled using the equation of motion (34a). On the other hand, replacement of the generalized momenta that appear in the term $\partial_x\pi(x)\partial_x\pi(x)$ of the equivalent Hamiltonian (32) using (34) might appear as an additional problem in the construction of the
  Lagrangian. We see that the problem is related to the particular form of the Hamiltonian (32) and does not appear in the equivalent Hamiltonian (31).

  We need the Lagrangian equations $-\frac{d}{dt}\frac{\delta L}{\delta\dot\phi}+\frac{\delta L}{\delta\phi}=0$ with the Lagrangian (35).
  Variation of the functional variation $\delta L$ due to a variation of $\delta\dot\phi$ gives
  \begin{equation}
  \delta_{\delta \dot\phi} L=\int dx \frac{\hbar}{2} \left[ \delta\dot\phi((V-\frac{\hbar^2}{2m}\Delta)^{-1}\dot\phi)+
  \dot\phi((V-\frac{\hbar^2}{2m}\Delta)^{-1}\delta\dot\phi)\right ]
  \end{equation}
  However, since the inverse of a Hermitian Hamiltonian $\hat H$ is also Hermitian the two terms in the previous expression are equal and the
   functional derivative  of $L$ with respect to $\dot\phi$ becomes
   \begin{equation}
\frac{\delta L}{\delta\dot\phi}=\hbar(V-\frac{\hbar^2}{2m}\Delta)^{-1}\dot\phi.
\end{equation}
The Lagrange equations become
\begin{equation}
\hbar^2(V-\frac{\hbar^2}{2m}\Delta)^{-1}\ddot\phi=-(V-\frac{\hbar^2}{2m}\Delta)\phi
\end{equation}
or
\begin{equation}
\hbar^2\ddot\phi=-(V-\frac{\hbar^2}{2m}\Delta)^2\phi.
\end{equation}
Of course, the Lagrange equation (39) is the special case of the general formula (19), as was the Lagrangian (35) the special case of (13).
No additional problems in the construction of the Lagrangian formulation for the Schr\"odinger equation (30) occur because of the special choice of the coordinate representation. Like in the general case, the only possible obstacle to the construction might be non-existence of the inverse operator
 $ (\hat H_R)^{-1}$ which is in this case $(V-\frac{\hbar^2}{2m}\Delta)^{-1}$.

\subsection{Klein-Gordon equation}

Consider the Schr\"odinger equation for a free relativistic one-dimensional scalar particle
\begin{equation}
i\hbar \frac{\partial \psi}{\partial t}=\hat H\psi=[\hbar^2c^2\Delta+m^2c^4]^{1/2}\psi.
\end{equation}

 Applying  the general formulas from section 2, or replacing everywhere in the previous example $-\frac{\hbar^2}{2m}\Delta +V(x)$ with $[\hbar^2c^2\Delta+m^2c^4]^{1/2}$   we arrive at the corresponding Lagrangian
 \begin{equation}
 L=\int dx \left[\frac{1}{2\hbar}\phi [\hbar^2c^2\Delta+m^2c^4]^{1/2} \phi+\frac{\hbar}{2}\dot\phi[(\hbar^2c^2\Delta+m^2c^4)^{1/2}]^{-1}\dot \phi\right],
 \end{equation}
 and the corresponding Lagrangian equations for real functions $\phi(t,x)$
 \begin{equation}
 \frac{1}{c^2}\frac{\partial^2 \phi}{\partial t^2}=\Delta\phi-(\frac{mc}{\hbar})^2\phi
 \end{equation}
 This is the Klein-Gordon equation for the real field $\phi(t,x)$.

\section{Summary}

In summary, we have developed the Lagrangian formalism for the abstract linear Schr\"odinger equation on a complex Hilbert space of a quantum system.
For a given Hamiltonian operator $\hat H=\hat H^R+i\hat H^I$ the Lagrangian system is expressed in terms of the operators $\hat H^R,\>\hat H^I$ and
 $(\hat H^R)^{-1}$, and is given by a second order equation on a real space. The operator $(\hat H^R)^{-1}$, which is crucial for construction of
  the Lagrangian formulation,
 exists provided that the spectrum of $\hat H$ is bounded away from zero. If this is not the case, than the Schr\"odinger equation is in fact equivalent to a constrained Hamiltonian system, with the corresponding singular Lagrangian formulation.
  A simple example illustrating the failure of the procedure that might occur in a singularly chosen basis is provided. The general formulation of the Lagrangian system is also illustrated in
  the eigenbasis of a Hamiltonian with a discrete spectrum and in the coordinate representation. The Klein-Gordon equation is seen as the Lagrangian
 system corresponding to the Schr\" odinger equation of a relativistic free particle.

{\bf Acknowledgments}

 This work was supported in part by the Ministry
  of  Education and Science of the Republic of Serbia, under project No.
  171017, 171028 and 171006.
and by COST (Action MP1006).

\end{document}